\documentclass[sigconf,  
]{acmart}

\usepackage{multicol}
\usepackage{booktabs} 
\usepackage{wrapfig}
\usepackage{graphicx}
\usepackage{subcaption} 

\setcopyright{none}

\acmConference[WWW'18]{The Web Conference}{April 2018}{Lyon, FR} 

\begin{document}
\title{Analyzing the Digital Traces of Political Manipulation: \\ The 2016 Russian Interference Twitter Campaign}


\author{Adam Badawy}
\affiliation{%
  \institution{USC Dept. of Political Science \& USC Information Sciences
Institute}
}
\email{abadawy@usc.edu}

\author{Emilio Ferrara}
\affiliation{%
  \institution{USC Information Sciences Institute}
}
\email{emiliofe@usc.edu}

\author{Kristina Lerman}
\affiliation{%
  \institution{USC Information Sciences Institute}
}
\email{lerman@isi.edu}

\begin{abstract}
Until recently, social media was seen to promote democratic discourse on social and political issues. However, this powerful communication platform has come under scrutiny for allowing hostile actors to exploit online discussions in an attempt to manipulate public opinion.  A case in point is the  ongoing U.S. Congress investigation of Russian interference in the 2016 U.S. election campaign, with Russia accused of, among other things, using trolls (malicious accounts created for the purpose of manipulation) and bots (automated accounts) to spread  misinformation and politically biased information. In this study, we explore the effects of this manipulation campaign, taking a closer look at users who re-shared the posts produced on Twitter by the Russian troll accounts publicly disclosed by U.S. Congress investigation. We collected a dataset with over 43 million elections-related posts shared on Twitter between September 16 and October 21, 2016 by about 5.7 million distinct users. This dataset included accounts associated with the identified Russian trolls. We use label propagation to infer the ideology of all users based on the news sources they shared. This method enables us to classify a large number of users as liberal or conservative with precision and recall above 90\%. 
Conservatives  retweeted Russian trolls about 31 times more often than liberals and produced 36 times more tweets. Additionally, most retweets of troll content originated from two Southern states: Tennessee and Texas. Using state-of-the-art bot detection techniques, we estimated that about 4.9\%  and 6.2\% of liberal and conservative users respectively were bots. 
Text analysis on the content shared by trolls reveals that they had a mostly conservative,  pro-Trump agenda.
Although an ideologically broad swath of Twitter users were exposed to Russian Trolls in  the period leading up to the 2016 U.S. Presidential election, it was mainly conservatives who helped amplify their message.

\end{abstract}


\maketitle

\section{Introduction}
Social media have helped foster democratic conversation about social and political issues: from the Arab Spring \cite{gonzalez2011dynamics}, to Occupy Wall Street movements \cite{conover2013digital, conover2013geospatial} and  other civil protests \cite{gonzalez2013broadcasters,varol2014evolution}, Twitter and other social media platforms appeared to play an instrumental role in involving the public in policy and political conversations by collectively framing the narratives related to particular social issues, and coordinating online and off-line activities. The use of digital media for political discussions  during presidential elections was examined by many  studies, including the past four U.S. Presidential elections \cite{adamic2005political, diakopoulos2010characterizing, bekafigo2013tweets, carlisle2013social, digrazia2013more}, and other countries like Australia \cite{gibson2006does, bruns2011use}, and Norway \cite{enli2013personalized}. Findings that focused on the positive effects of social media, such as increasing voter turnout \cite{bond201261}  or exposure to diverse political views \cite{bakshy2015exposure} contributed to the general praise of these platforms as a tool for promoting democracy and civic engagement \cite{shirky2011political, loader2011networking, effing2011social, tufekci2012social, tufekci2014big}. 

However, concerns regarding the possibility of manipulating public opinion and spreading political misinformation or fake news through social media were also raised early on \cite{howard2006new}. These effects were later documented by several studies \cite{ratkiewicz2011detecting, conover2011political, el2013social, woolley2016automation, shorey2016automation, bessi2016social, ferrara2017disinformation, fourney2017geographic}. Social media have been proven as effective tools to influence individuals' opinions and behaviors \cite{aral2009distinguishing, aral2012identifying,bakshy2011everyone, centola2011experimental,centola2010spread} and some studies even evaluated the current tools to combat misinformation \cite{Pennycook2017b}. 
Computational tools, like troll accounts and social bots, have been designed to perform such type of influence operations at scale, by cloning or emulating the activity of human users while operating at much higher pace (e.g., automatically producing content following a scripted agenda) \cite{hwang2012socialbots, messias2013you, ferrara2016rise, varol2016online} -- however, it should be noted that bots have been also used, in some instances, for positive interventions \cite{savage2016botivist, monsted2017evidence}.

Early accounts of the adoption of bots to attempt manipulate political communication with misinformation started in 2010, during the U.S. midterm elections, when social bots were employed to support some candidates and smear others; in that instance, bots injected thousands of tweets pointing to Web sites with fake news \cite{ratkiewicz2011truthy}. Similar cases were reported during the 2010 Massachusetts special election \cite{metaxas2012social} -- these campaigns are often referred as to Twitter bombs, or political astroturf. Unfortunately, oftentimes determining the actors behind these operations was impossible \cite{kollanyi2016bots, ferrara2016rise}. Prior to this work, only a handful of other operations were linked to some specific actors \cite{woolley2016automation}, e.g., the alt-right attempt to smear a presidential candidate before the 2017 French election \cite{ferrara2017disinformation}. This is because governments, organizations, and other entities with sufficient resources, can obtain the technological capabilities necessary to covertly deploy hundreds or thousands of accounts and use them to either support or attack a given political target. Reverse-engineering these strategies has proven a challenging research venue \cite{freitas2015reverse, alarifi2016twitter, subrahmanian2016darpa, davis2016botornot}, but it can ultimately lead to techniques to identify the actors behind these operations.

Manipulation through misinformation, or ``fake news,'' has in the past year gain notoriety,  as a result of the 2016 U.S. presidential election~ \cite{allcott2017social, shao2017spread, pennycook2017falls, mele2017combating, guess2018selective, zannettou2018disinformation}. Data from Facebook and Twitter show that deceptive, made-up content, marketed as political news, was shared with millions of Americans before the 2016 election,\footnote{\url{https://blog.twitter.com/official/en_us/topics/company/2017/Update-Russian-Interference-in-2016--Election-Bots-and-Misinformation.html}}$^,$\footnote{\url{https://www.theguardian.com/technology/2017/oct/30/facebook-russia-fake-accounts-126-million}} although only a handful of studies have examined this phenomenon in detail \cite{guess2018selective}.
One difficulty facing such studies is objectively determining what is fake news, as there is a range of untruthfulness from simple exaggeration to outright lies. Beyond factually wrong information, it is difficult to classify information as fake. We argue that the key element in the definition of fake news is intent. In order to label some accounts or sources of information as fake, an \textit{intent to deceive} has to be present. A malicious intent to harm the political process and cause distrust in the political system was evident in 2,752 now-deactivated Twitter accounts that were later identified as being tied to Russia's ``Internet Research Agency" troll farm. The U.S. Congress released a list of these accounts as part of the Congress' investigation of Russian efforts to interfere in the 2016 presidential election. 
Since their intent was clearly malicious, we use messages posted by these Russian Troll accounts as a proxy for fake news, and study their spread on Twitter to understand the phenomenon of misinformation and its effects on the modern political life.

In this paper, we aim to answer three crucial  questions regarding the spread of misinformation. 
The first question is: Does political ideology affect who engages with the producers of fake news and helps propagate it?  Is the fake news phenomenon more pronounced among liberals or conservatives, or is it evenly spread across the political spectrum? Second, how active were bots in spreading fake news, and where on the political spectrum was this phenomenon more prevalent? Last, did the fake news phenomenon have geographical component, with users located within some states participating in the consumption and propagation of misinformation more than others?

We collected Twitter data over a  period of few weeks in the months leading up to the election. By continuously polling the Twitter Search API for relevant, election-related content using hashtag- and keyword-based queries, we obtained a  dataset of over 43 million tweets generated by about 5.7 million distinct users between September 16 and October 21, 2016.
We were able to successfully determine the political ideology of most of the users using label propagation on the retweet network  with precision and recall exceeding 90\%. Next, using advanced machine learning techniques developed to discover social bots \cite{ferrara2016rise, subrahmanian2016darpa, davis2016botornot} on users who engaged with Russian trolls, we found that bots existed among both liberal and conservative  users (although it is worthy to note that most of these users are conservative and pro-Trump). We performed text analysis on the content Russian trolls disseminated, and found that they were mostly concerned with conservative causes and were spreading pro-Trump material. Additionally, we offer an extensive geospatial analysis of tweets across the United States, showing that it is, as expected, proportionate to the states' population size. 

\subsection{Research Questions}

Our work attempts to characterize users who engaged with Russian trolls by resharing their messages in the period leading to the 2016 US Presidential election. These interactions serve as a proxy for fake news consumption. Specifically, we examine the following questions:

\begin{itemize}
\item What was the role of the users' political ideology?
\item How many of these accounts were bots, and how was bot activity distributed among liberal and conservative sides? 
\item What was the geographic location of the users who engaged with Russian Trolls? Did trolls especially succeed in specific areas of the US?
\end{itemize}

\subsection{Summary of Contributions}

Our findings  can be summarized as follows:

\begin{itemize}
\item  We proposed a novel way of measuring the production and consumption of fake news through the analysis of activities of manipulative accounts (Russian Trolls) on Twitter in the run-up period to the election. 
\item Using a network-based machine learning method, we were able to successfully determine the political ideology of most of the users in our dataset, with precision and recall above 90\%.
\item We ran state-of-the-art bot detection analysis on users who engaged with Russian Trolls and determined that bots were engaged in both liberal and conservative domains (with the caveat that the majority of the users in our dataset are conservative; thus, most bots were on the conservative side as well).
\item Text analysis shows that Russian trolls were mostly promoting conservative causes and were, specifically, spreading pro-Trump material. 
\item We offered a comprehensive geo-spatial analysis of the tweets.  
\end{itemize}
Our comprehensive analysis indicates that although the consumption and dissemination of content produced by  Russian Trolls was distributed broadly over the political spectrum, it was especially concentrated among the conservative Twitter accounts. These accounts helped amplify the misinformation produced by trolls to manipulate public opinion during the period leading up to the 2016 U.S. Presidential election.

\section{Data Collection}
\subsection{Twitter Dataset}

We created a list of hashtags and keywords that relate to the 2016 U.S. Presidential election. The list was crafted to contain a roughly equal number of hashtags and keywords associated with each major Presidential candidate: we selected 23 terms, including five terms referring to the Republican Party nominee Donald J. Trump (\#donaldtrump, \#trump2016, \#neverhillary, \#trumppence16, \#trump), four terms for Democratic Party nominee Hillary Clinton (\#hillaryclinton, \#imwithher, \#nevertrump, \#hillary), and several terms related to debates. To make sure our query list was comprehensive, we also added a few keywords for the two third party candidates, including the Libertarian Party nominee Gary Johnson (one term), and Green Party nominee Jill Stein (two terms). 

\begin{figure}
  \includegraphics[width=\linewidth]{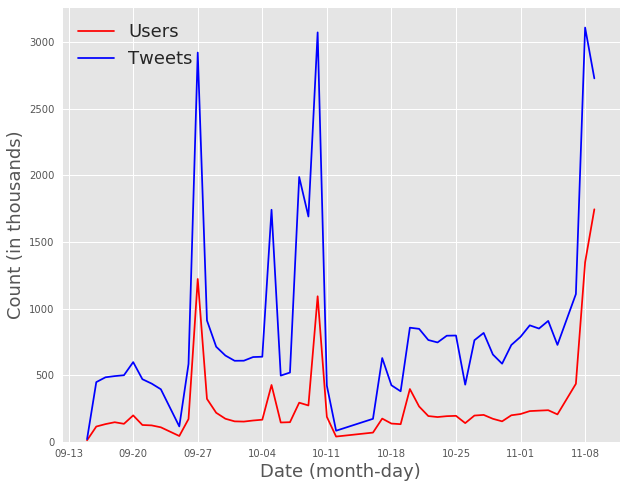}
  \caption{Timeline of the volume of tweets (in blue) and users (in red) generated during our observation period.}
  \label{Figure 1}
\end{figure}

By querying the Twitter Search API at an interval of 10 seconds, continuously and without interruptions between 15\textsuperscript{th} of September and 9\textsuperscript{th} of November 2016, we collected a large dataset containing 43.7 million unique tweets posted by nearly 5.7 million distinct users. Table \ref{Table 1} reports some aggregate statistics of the dataset while Figure \ref{Figure 1} shows the timeline of the volume of the tweets and users during the aforementioned period. The data collection infrastructure ran inside an Amazon Web Services (AWS) instance to ensure resilience and scalability. We chose to use the Twitter Search API to make sure that we obtained all tweets that contain the search terms of interest posted during the data collection period, rather than a sample of unfiltered tweets. This precaution we took avoids certain issues related to collecting sampled data using the Twitter Stream API that had been reported in literature \cite{morstatter2013sample}.

\begin{table}[tbp]
\centering
\caption{Twitter Data Descriptive Statistics.}
\label{Table 1}
\begin{tabular}{ll}
Statstic                         & Count      \\ \midrule
\# of Tweets                     & 43,705,293 \\
\# of Retweets                   & 31,191,653 \\
\# of Distinct Users              & 5,746,997  \\ 
\# of Tweets/Retweets with a Url & 22,647,507
\end{tabular}
\end{table}

\subsection{Classification of Media Outlets}
We classify users by their ideology based on the political leaning of the media outlets they shared. The classification algorithm is described later in the paper; here, we describe the methodology of obtaining ground truth labels for these outlets.


We use lists of partisan media outlets compiled by third-party organizations, such as AllSides\footnote{\url{https://www.allsides.com/media-bias/media-bias-ratings}} and Media Bias/Fact Check.\footnote{\url{https://mediabiasfactcheck.com/}} The combined list includes 249 liberal outlets and 212 conservative outlets. After cross-referencing with  domains obtained in our  Twitter dataset, we identified 190 liberal and 167 conservative outlets. We picked five media outlets from each partisan category that appeared most frequently in our Twitter dataset and compiled a list of users who tweeted from these outlets. The list of media outlets/domain names for each partisan category is reported in Table \ref{Table 2}.

\begin{figure}
  \begin{subfigure}[b]{0.4\textwidth}
    \includegraphics[width=\textwidth]{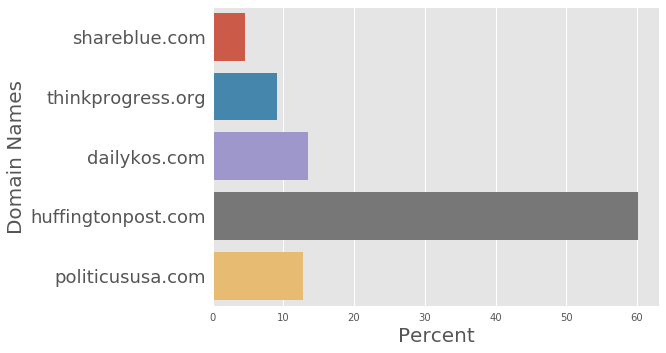}
    \caption{}
    \label{figure 2}
  \end{subfigure}
  \begin{subfigure}[b]{0.4\textwidth}
    \includegraphics[width=\textwidth]{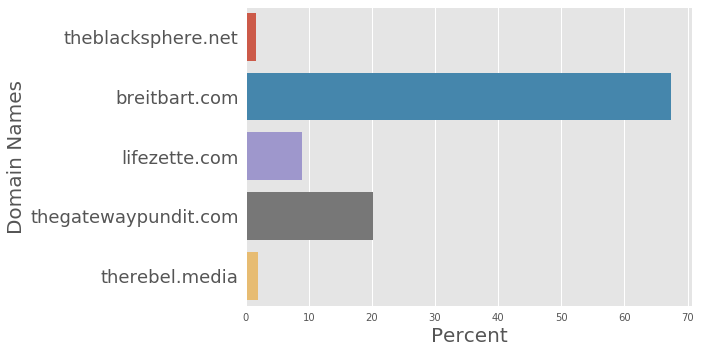}
    \caption{}
    \label{figure 3}
  \end{subfigure}
  \caption{Distribution of tweets with links to the top five (a) liberal and (b) conservative media outlets.}
\end{figure}

Overall, 161,907  tweets in the dataset contained a url that pointed to one of the top-five liberal media outlets, which were tweeted by 10,636 users. For the conservative outlets, the numbers are 184,720 tweets and 7,082 users. Figures \ref{figure 2} and \ref{figure 3} show the distribution of tweets with urls from liberal and conservative outlets respectively. As we can see in the figures, Huffington Post and Breitbart make up more than 60\% of the total volume. 

We used a polarity rule to label Twitter users as liberal or conservative depending on the number of tweets they produced with links to liberal or conservative sources. In other words, if a user had more tweets with urls to liberal sources, he/she would be labeled liberal and vice versa. Although the overwhelming majority of users include urls that are either liberal or conservative, we removed any users that had equal number of tweets from each side \footnote{We used five categories, as in left, left center, center, right center, right, to make sure we have a final list of users who are unequivocally liberal or conservative and do not fall in the middle. The media outlet lists for the left/right center and center were compiled from the same sources.}. Our final set of labeled users include 29,832 users.

\begin{table}[tbp]
\centering
\caption{Liberal \& Conservative Domain Names.}
\label{Table 2}
\begin{tabular}{@{}ll@{}}
\toprule
Liberal                & Conservative             \\ \midrule
www.huffingtonpost.com & www.breitbart.com        \\
thinkprogress.org      & www.thegatewaypundit.com \\
www.politicususa.com   & www.lifezette.com        \\
shareblue.com          & www.therebel.media       \\
www.dailykos.com       & theblacksphere.net       \\ \bottomrule
\end{tabular}
\end{table}

\subsection{Russian Trolls}

We used a list of 2,752 Twitter accounts identified as Russian trolls that was compiled and released by the U.S. Congress\footnote{See https://www.recode.net/2017/11/2/16598312/russia-twitter-trump-twitter-deactivated-handle-list}, see Table \ref{Table 3} for descriptive statistics. Out of the accounts appearing on the list, 221  exist in our Twitter dataset, and 85 of them wrote original tweets (861 tweets). Russian trolls in our dataset retweeted 2,354 other distinct users 6,457 times. Trolls retweeted each other only 51 times. 

Twitter users can choose to report their location in their profile. Most of the self-reported locations of accounts associated with Russian trolls were within the U.S. (some provided Russian locations in their profile), and most of the tweets were from users who are based in Tennessee and Texas, 49,277 and 26,489 respectively. 


Russian trolls were retweeted 83,719 times, but most of these retweets were for three troll accounts only: \lq{TEN\_GOP}\rq, 49,286; \lq{Pamela\_Moore13}\rq, 16,532; and \lq{TheFoundingSon}\rq, 8,755, in total making over 89\% of the times Russian trolls were retweeted.  Russian trolls were retweeted by 40,224 distinct users.

\begin{table}[]
\centering
\caption{Descriptive Statistics on Russian Trolls.}
\label{Table 3}
\begin{tabular}{@{}ll@{}}
\toprule
  & Value                                    \\ \midrule
\# of Russian Trolls               & 2,735  \\
\# of trolls in our data           & 221    \\
\# of trolls wrote original tweets & 85     \\
\# of original tweets              & 861    \\
\bottomrule
\end{tabular}
\end{table}

\section{Data Analysis \& Methods}
\subsection{Retweet Network}
We construct a retweet network, containing nodes (Twitter users) with a directed link between them if one user retweeted a post of another. 
Table \ref{Table 4} shows the descriptive statistics of the retweet network. It is a sparse network with a giant component that includes 4,474,044 nodes.


\begin{table}[tbp]
\centering
\caption{Descriptive statistics of the Retweet Network.}
\label{Table 4}
\begin{tabular}{@{}ll@{}}
\toprule
Statstic       & Count      \\ \midrule
\# of nodes    & 4,678,265  \\
\# of edges    & 19,240,265 \\
Max in-degree  & 278,837    \\
Max out-degree & 12,780     \\
Density        & 8.79E-07   \\ \bottomrule
\end{tabular}
\end{table}


\subsection{Label Propagation}

We used label propagation\footnote{We used the algorithm in the Python version of the Igraph library \cite{csardi2006igraph}} to classify Twitter accounts as liberal or conservative. 
In a network-based label propagation algorithm each node is assigned a label, which is updated iteratively based on the labels of node's network neighbors. In  label propagation, a node takes the most frequent label of its neighbors as its own new label. The algorithm proceeds updating labels iteratively and stops when 
the labels no longer change (see \citep{raghavan2007} for more information). The algorithm takes as parameters (i) weights, in-degree or how many times node $i$ retweeted node $j$;  (ii) seeds (the list of labeled nodes). We fix the seeds' labels so they do not change in the process, since this seed list also serves as our ground truth. 

We constructed a retweet network where each node corresponds to a Twitter account and a link exists between pairs of nodes when one of them retweets a message posted by the other. 
We used the 29k users mentioned in the media outlets sections as seeds, those who mainly retweet messages from either the liberal or the conservative media outlets in table \ref{Table 2}, and label them accordingly. We then run label propagation to label the remaining nodes in the retweet network.

To validate results of the label propagation algorithm, we applied stratified cross (5-fold) validation to the set of 29k seeds. We train the algorithm on 4/5 of the seed list and see how it performs on the remaining 1/5. The precision and recall scores are around 0.91.

To further validate the labeling algorithm, we noticed that a group of twitter accounts  puts media outlet urls as their personal link/website. We compiled a list of these hyper-partisan  twitter users who has the domain names from table \ref{Table 2} in the profiles and used the same approach explained in the previous paragraph (stratified 5-fold cross-validation). The precision and recall scores for the test set for these users were around 0.93. Table \ref{Table 5} show the precision and recall scores for the two validation methods we used, both labeled  more than 90\% of the test set users correctly, cementing our confidence in the performance of the labeling algorithm.

\begin{table}[]
\centering
\caption{Precision \& Recall scores for the seed users and hyper-partisan users test sets.}
\label{Table 5}
\begin{tabular}{@{}lll@{}}
\toprule
 & Seed Users & Hyper-Partisan Users       \\ \midrule
Precision  & 0.91                 & 0.93 \\
Recall     & 0.91                 & 0.93 \\ \bottomrule
\end{tabular}
\end{table}

\subsection{Bot Detection}

Determining whether either human or a bot controls a social media account has proven a very challenging task \cite{ferrara2016rise, subrahmanian2016darpa}. We used an openly accessible solution called Botometer (a.k.a. BotOrNot) \cite{davis2016botornot}, consisting of both a public Web site (\url{https://botometer.iuni.iu.edu/})  and a Python API (\url{https://github.com/IUNetSci/botometer-python}), which allow for making this determination. Botometer is a machine-learning framework that extracts and analyses a set of over one thousand features, spanning content and network structure, temporal activity, user profile data, and sentiment analysis to produce a score that suggests the likelihood that the inspected account is indeed a social bot. Extensive analysis revealed that the two most important classes of features to detect bots are, maybe unsurprisingly, the metadata and usage statistics associated with the user accounts. The following indicators provide the strongest signals to separate bots from humans: (i) whether the public Twitter profile looks like the default one or it is customized (it requires some human efforts to customize the profile, therefore bots are more likely to exhibit the default profile setting); (ii) absence of geographical metadata (humans often use smartphones and the Twitter iPhone/Android App, which records as digital footprint the physical location of the mobile device); and, (iii) activity statistics such as total number of tweets and frequency of posting (bots exhibit incessant activity and excessive amounts of tweets), proportion of retweets over original tweets (bots retweet contents much more frequently than generating new tweets), proportion of followers over followees (bots usually have less followers and more followees), account creation date (bots are more likely to have recently-created accounts), randomness of the username (bots are likely to have randomly-generated usernames). 

Botometer was trained with thousands of instances of social bots, from simple to sophisticated, with an accuracy above 95 percent~\cite{davis2016botornot}. Typically, Botometer yields likelihood scores above 50 percent only for accounts that look suspicious to a scrupulous analysis. We adopted the Python Botometer API to systematically inspect the most active users in our dataset. The Python Botometer API queries the Twitter API to extract 300 recent tweets and publicly available account metadata, and feeds these features to an ensemble of machine learning classifiers, which produce a bot score.

To label accounts as bots, we use the fifty-percent threshold -- which has proven effective in prior studies \cite{davis2016botornot} -- an account is considered to be a bot if the bot score is above 0.5.

\subsection{Geo-location}

There are two ways to identity the location of tweets produced by users. One way is to collect the  coordinates of the location the tweets were sent from; however, this is only possible if users enable the geolocation option on their Twitter accounts. The second way is to analyze the self-reported home locations in  users' profiles. The latter includes substantially more noise, since many people write fictitious or imprecise locations, for example, they may identify the state and the country they reside in, but not the city.

There were 36,351 tweets with exact coordinates in our dataset. The distribution of tweets across the fifty states tended to be concentrated in the South, with Kentucky being the state with the highest number of geolocated tweets. It is hard to know why that is the case; besides, geo-tagged tweets in this dataset comprise less than 0.001\% of the whole dataset.

\begin{figure}
  \includegraphics[width=\linewidth,clip=true,trim=0 20 0 30]{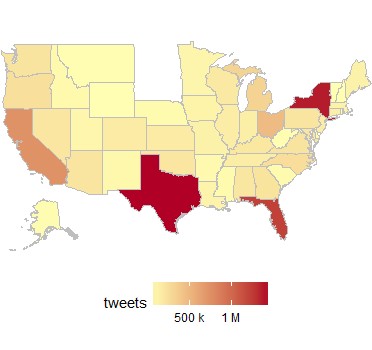}
  \caption{Self-reported sources for tweets; white/non-existing states mean no tweets/users are located within these states.} 
  \label{figure4}
\end{figure}

Tweets and users' self-reported locations make up substantially more of our dataset than geo-tagged tweets. More than 3.8 million Twitter users provided a location in their profile, and out of those that are intelligible and located within the US, 1.6 Million remained. From users' locations, we mapped over 10.5 Million tweets to some U.S, States, as shown in Figure \ref{figure4}. The distribution of the tweets and users seems to be as expected population-wise, although it is slightly less than expected for the state of California, provided that it is the most populous state in the nation.



\section{Results}

\subsection{Activity of Russian Trolls}

Although the predicted labels for the 215 Russian troll accounts in our dataset is almost equally divided between liberal and conservative, with 107 accounts labeled as liberal and 108 labeled as conservative, the two groups are extremely different in terms of their activity (see table \ref{Table 6}). While there are only 15 liberal Russian trolls who wrote original tweets, there are 64 conservative trolls who produced original content. Left leaning trolls wrote 44 original tweets, while conservatives wrote 844 original tweets. Table \ref{Table 7} shows the top 20 stem words from tweets of liberal and conservative trolls respectively. 

\begin{table}[]
\centering
\caption{Breakdown of the Russian Trolls by political ideology, with the ratio of conservative to liberal trolls.}
\label{Table 6}
\begin{tabular}{@{}llll@{}}
\toprule
& Liberal                        & Conservative & Ratio     \\ \midrule
\# of trolls                   & 107          & 108         & 1   \\
\# of trolls w orginial tweets & 15           & 64          & 4.3 \\
\# of original tweets          & 44           & 844         & 19  \\ \bottomrule
\end{tabular}
\end{table}


\begin{table}[]
\centering
\caption{Top 20 stemmed words from the tweets of Russian Trolls classified as liberal and conservative.}
\label{Table 7}
\begin{tabular}{@{}llll@{}}
\toprule
Liberal  & count & Conservative & count \\ \midrule
trump & 14 & trumpforpresid & 486 \\
debat & 10 & trump & 241 \\
nevertrump & 6 & trumppence16 & 227 \\
like & 5 & hillaryforprison2016 & 168 \\
2016electionin3word & 5 & vote & 127 \\
elections2016 & 4 & maga & 113 \\
imwithh & 4 & neverhillari & 106 \\
obama & 3 & election2016 & 102 \\
need & 3 & hillari & 100 \\
betteralternativetodeb & 3 & hillaryclinton & 85 \\
women & 3 & trump2016 & 80 \\
would & 3 & draintheswamp & 50 \\
vote & 3 & trumptrain & 48 \\
mondaymotiv & 2 & debat & 48 \\
last & 2 & realdonaldtrump & 45 \\
oh & 2 & electionday & 43 \\
thing & 2 & clinton & 41 \\
damn & 2 & makeamericagreatagain & 34 \\
see & 2 & votetrump & 32 \\
defeat & 2 & america & 31 \\ \bottomrule
\end{tabular}
\end{table}



\subsection{Users Engaged with Russian Trolls}

Concerning the users who retweeted Russian trolls, which we call spreaders, three key questions emerge: What is their political ideology (liberal vs conservative)? Where are they located? How many of them are bots?

\subsubsection{Political Ideology}

Spreaders tell a fascinating story (see tables \ref{Table 8} \&  \ref{Table 9}). There are 28,274 spreaders in our dataset that wrote original tweets. They produced over 1.5 Million original tweets and over 12 Million tweets and retweets, not counting the ones from Russian trolls. Looking at the content of the top 10 users, we can easily identify them as conservative; besides, they produced an unreasonable amount of tweets in such a short period. In the next paragraph we will look systematically at these users' activities by political leaning. 

There are more than 42 thousand tweets by the liberal spreaders and more than 1.5 million tweets by conservative ones. There are 892 liberal  and 27,382 conservative spreaders. The top stemmed words in the liberals' tweets indicate support for Clinton, while the conservatives' postings openly support Trump. The top urls for the liberals include media outlets, such as: Huffington Post and NBC News, while conservatives tweeted from Breitbart, The Gateway Pundit, and Info Wars. For the profile url, liberals mostly had social network accounts, while conservatives, besides social network accounts, put ``www.donaldjtrump.com'' and ``lyingcrookedhillary.com''. 

\begin{table}[]
\centering
\caption{Descriptive statistics of spreaders, i.e., users who retweeted Russian Trolls.}
\label{Table 8}
\begin{tabular}{@{}ll@{}}
\toprule
  & Value                                                          \\ \midrule
\# of spreaders                      & 40,224                   \\
\# of times retweeted trolls    & 83,719                   \\
\# of spreaders with original tweets & 28,274                   \\
\# of original tweets                & \textgreater 1.5 Million \\
\# of original tweets and retweets   & \textgreater 12 Million  \\ \bottomrule
\end{tabular}
\end{table}

\begin{table}[]
\centering
\caption{Breakdown by political ideology of users who spread Russian Troll content and wrote original tweets.
}
\label{Table 9}
\begin{tabular}{@{}llll@{}}
\toprule
 &  Liberal         & Conservative        & Ratio                   \\ \midrule
\# of spreaders & 892                 & 27,382                   & 31 \\
\# of tweets    & \textgreater 42,000 & \textgreater 1.5 Million & 36 \\ \bottomrule
\end{tabular}
\end{table}

\subsubsection{Geospaitial Analysis}
\begin{figure}
  \begin{subfigure}[b]{0.4\textwidth}
    \includegraphics[width=\textwidth,clip=true,trim=0 20 0 20]{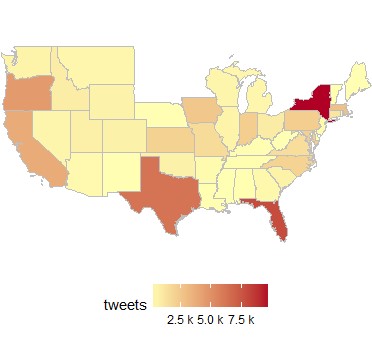}
    \caption{}
    \label{figure 5}
  \end{subfigure}
  \begin{subfigure}[b]{0.4\textwidth}
    \includegraphics[width=\textwidth,clip=true,trim=0 20 0 20]{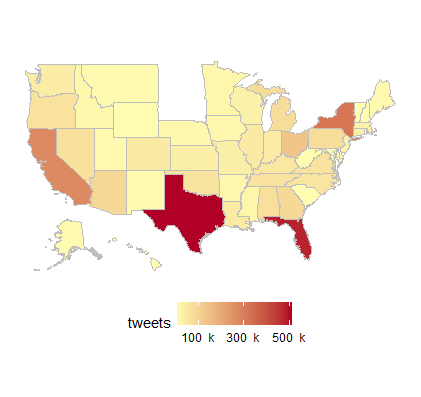}
    \caption{}
    \label{figure 6}
  \end{subfigure}
  \caption{Self-reported sources for tweets for liberal users who retweeted Russian trolls (a), and for conservatives (b); white/non-existing states mean no tweets/users are located within these states.}
\end{figure}

we can see in Figures \ref{figure 5}  and \ref{figure 6}, liberals' tweets come from fewer states and some Democratic states stand out as a major source of the tweets, such as the state of New York. For the conservative users, the tweets come from higher number of states (which can be just an artifact of the conservatives' tweet volume) and prominent Republican states, such as Texas and Florida stand out as the biggest geographic sources of the conservatives' tweets.

\subsubsection{Bots}
Using the approach explained in the Bot detection section, we were able to obtain bot scores for 34,160 out of the 40,224 spreaders. The number of accounts that has a bot score above 0.5 and can be considered bots are 2,126 accounts. 

Answering the third question is trickier, since most of the spreaders are conservative (see table \ref{Table 10} for spreaders' bot analysis by political ideology). But putting that aside, out of the 34,160 spreaders with bot scores, 1,506 are liberal and 75 of them have bot scores above 0.5, about 4.9\% of the total. As for the conservatives, there are 32,513 spreaders, with 2,018 who have bot scores more than 0.5, representing around 6.2\% of the total. In terms of tweet/retweet production, liberal spreaders  produced 224,943 tweets/retweets with 18,749 tweets/retweets by users who have a bot score above 0.5, representing around 8.3\%. For conservative spreaders, they produced 11,928,886 tweets/retweets, with 955,583 from users with bot score more than 0.5, around 8\% of the total.

\begin{table}[]
\centering
\caption{Bot Analysis on Spreaders (those with bot scores).}
\label{Table 10}
\begin{tabular}{@{}llll@{}}
\toprule
& Liberal              & Conservative & Ratio    \\ \midrule
\# of spreaders      & 1,506        & 32,513      & 22 \\
\# of tweets         & 224,943      & 11,928,886  & 53 \\
\# of bots           & 75           & 2,018       & 27 \\
\# of tweets by bots & 18,749       & 955,583     & 51 \\ \bottomrule
\end{tabular}
\end{table}

\begin{figure}
  \begin{subfigure}[a]{0.4\textwidth}
    \includegraphics[width=\textwidth]{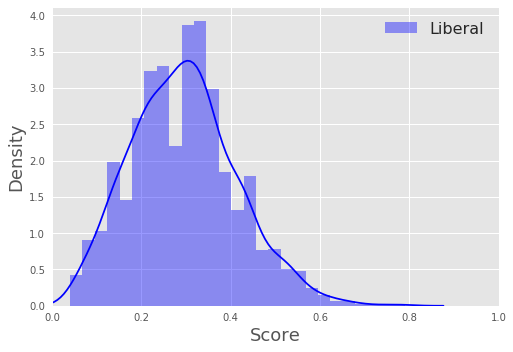}
  \end{subfigure}
  \begin{subfigure}[b]{0.4\textwidth}
    \includegraphics[width=\textwidth]{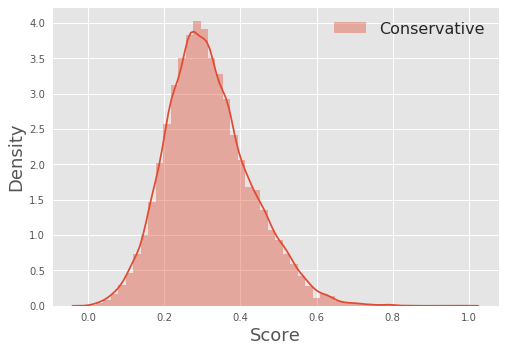}
  \end{subfigure}
  \caption{Distribution of the probability density of bot scores assigned to liberal users who retweet Russian Trolls (top) and for conservative users (bottom).}
  \label{fig5}
\end{figure}

Figure \ref{fig5} shows the probability density of bot scores of liberal (top) and conservative (bottom) spreaders respectively. Again, putting the disproportionate number of liberals to conservatives aside, the density of bot scores seem to be similar to each other, with the majority of the users ranging from 0 to 0.6.

\section{Conclusions}

The dissemination of information and the mechanisms for democratic discussion have radically changed since the advent of digital media, especially social media. Platforms like Twitter have been extensively praised for their contribution to democratization of public discourse on civic and political issues. However, many studies have also highlighted the perils associated with the abuse of these platforms. The spread of deceptive, false and misleading information aimed at manipulating public opinion are among those risks.

In this work, we investigated the role and effects of misinformation, using the content produced by Russian Trolls on Twitter as a proxy for misinformation. We collected tweets posted during the period between 16 September and 21 October 2016 related to the U.S. presidential election using the Twitter Search API and a manually compiled list of keywords and hashtags. 
We showed that that misinformation (produced by Russian Trolls) was shared more widely by conservatives than liberals on Twitter. Although there were about 4 times as many  Russian Trolls posting conservative views as liberal ones, the former produced almost 20 times more content. In terms of users who retweeted these trolls, there were about 30 times more conservatives than liberals. Conservatives also outproduced liberals in terms on content, at a rate of 35:1.  Using  state-of-the-art bot detection method, we estimated that about 4.9\%  and 6.2\% of the liberal and conservative users are bots. 

The spread of misinformation by malicious actors can have severe negative consequences. It can enhance malicious information and polarize political conversations, causing confusion and social instability. Political scientists are currently investigating the consequences of such phenomena \cite{woolley2016automation,shorey2016automation}. We plan to explore in particular the issue of how malicious information spread via exposure and the role of peer effect. 
Concluding, it is important to stress that, although our analysis unveiled the current state of the political debate and agenda pushed by the Russian Trolls who spread malicious information, it is impossible to account of all the malicious efforts aimed at manipulation during the last presidential election. State- and non-state actors, local and foreign governments, political parties, private organizations, and even individuals with adequate resources~\cite{kollanyi2016bots}, could obtain operational capabilities and technical tools to construct misinformation campaigns and deploy armies of social bots to affect the directions of online conversations. Therefore, future efforts will be required by the machine learning research and social sciences communities to study this issue in depth and develop more sophisticated detection techniques capable of unmasking and fighting these malicious efforts.

\begin{acks}
\small{The authors gratefully acknowledge support by the Air Force Office of Scientific Research (AFOSR, award number FA9550-17-1-0327). The views and conclusions contained herein are those of the authors and should not be interpreted as necessarily representing the official policies or endorsements, either expressed or implied, of AFOSR or the U.S. Government.}
\end{acks}

\bibliographystyle{ACM-Reference-Format}
\bibliography{citations.bib}
\end{document}